\documentclass[12pt,preprint]{aastex}

\def\deg{\ifmmode {^{\circ}}\else {$^\circ$}\fi}

\slugcomment{Submitted to the {\it Astrophysical Journal Letters}}
\shorttitle{{\it Spitzer} Constraints on Lensed Galaxy}
\shortauthors{Chary, Stern \& Eisenhardt}

\begin{document}

\title{{\it Spitzer} Constraints on the ${\mathbf z=6.56}$ Galaxy Lensed
by Abell~370}

\author{Ranga-Ram Chary\altaffilmark{1}, Daniel Stern\altaffilmark{2}
\& Peter Eisenhardt\altaffilmark{2}}

\altaffiltext{1}{{\it Spitzer} Science Center, California Institute of
Technology, Pasadena, CA 91125; {\tt rchary@caltech.edu}}
\altaffiltext{2}{Jet Propulsion Laboratory, California Institute of
Technology, Pasadena, CA 91109}

\begin{abstract}

We report on {\it Spitzer} IRAC observations of the spectroscopically
confirmed $z=6.56$ lensed Ly$\alpha$ emitting source HCM~6A which was found behind
the cluster Abell~370.  Detection of the source at 3.6 and 4.5~$\mu$m, corresponding
to rest-frame optical emission, allows us to study the stellar population of this
primeval galaxy. The broadband flux density at
4.5~$\mu$m is enhanced compared to the continuum at other
wavelengths, likely due to the presence of strong
H$\alpha$ in emission. The derived H$\alpha$ line flux corresponds to a
star-formation rate of $\gtrsim 140$~M$_{\sun}$~yr$^{-1}$, 
more than an order of magnitude larger than estimates from the ultraviolet
continuum and Ly$\alpha$ emission line. The dust extinction
required to explain the discrepancy is $A_{\rm V} \sim 1$~mag. 
The inference of dust at such high redshifts is surprising and implies
that the first epoch of star-formation in this galaxy occurred
at $z\sim20$. 

\end{abstract}


\section{Introduction}

The end of the ``Dark Ages'', sometime between redshifts of 6$-$50,
is signalled by the formation of the first galactic and proto-galactic structures.
The detection and characterization of these early objects is essential for identifying
the sources of re-ionization and revealing if galaxies in the early Universe
build their stellar mass primarily through monolithic collapse or hierarchical merging. 
Potential sources responsible for re-ionization include supermassive 
black holes in
bright QSOs, violently star-forming galaxies and exotic but hypothetical
Population~III stars thought to be of mass $>100$~M$_{\sun}$.  However,
QSOs have been effectively ruled out based on their luminosity
function at $z\sim6$ determined from the Sloan Digitized Sky
Survey \citep{Fan:02} and deeper, narrower field of view quasar
surveys \citep{Willott:05, Mahabal:05}. Star-forming galaxies are
likely to dominate the re-ionization only if the faint end slope of
the UV luminosity function at $z \sim 6$ is steeper than the slope at
$z \sim 3$ \citep{Yan:04}.  However, of all the $z > 6$ candidates
identified in various survey fields \citep[e.g.,][]{Bouwens:04,
Dickinson:04, Stanway:05}, the only spectroscopically confirmed $z >
6$ sources are the emission line galaxies \citep[e.g.,][]{Rhoads:04,
Stern:05a, Taniguchi:05}.  The first spectroscopically confirmed
galaxy at $z > 6$ was HCM~6A ($z = 6.56$), discovered by \citet{Hu:02}
in a narrowband imaging survey of lensing galaxy cluster fields and
confirmed through the detection of redshifted Ly$\alpha$.  The source
is magnified by the foreground cluster Abell~370 by a factor of 4.5 and
appears to be fragmented into two components separated by $<2\arcsec$
in the North-East/South-West direction. The Ly$\alpha$ line and UV
continuum luminosity correspond to a star-formation rate (SFR) of $2-9$
M$_{\sun}$ yr$^{-1}$. This does not include any corrections for dust,
since the existence of dust and absorption are quite uncertain from the
purely rest-frame ultraviolet data previously obtained for HCM~6A.

In this {\em Letter} we report on the {\it Spitzer} detection of HCM~6A
using the Infrared Array Camera \citep[IRAC;][]{Fazio:04a}.  IRAC,
due to its exquisite sensitivity, has been successful in detecting
the rest-frame optical emission from
candidate and confirmed $z>5$ sources in the GOODS fields
\citep[e.g.,][]{Eyles:05, Mobasher:05, Yan:05} and in fields which are
lensed by massive clusters \citep{Egami:05}.  
These measurements provide a better
constraint on the physical properties of high redshift galaxies
such as the stellar mass, age of the
stellar population and internal dust extinction.

\section{Observations and Reductions}

The {\it Spitzer}/IRAC observations were undertaken as part of
guaranteed time observations. 
Data at 3.6~$\mu$m and 5.8~$\mu$m were taken
with 12 dithered 200~sec exposures, for a total on-source integration time
of 2400~sec. At 4.5~$\mu$m and 8.0~$\mu$m, 18 dithered 200~sec exposures were used for a
total on-source integration time of 3600~s.  

We made minor modifications to the pipeline-processed data.  Beginning
with the basic calibrated data from the IRAC pipeline (version
S11.0.2), we corrected the ``pulldown'' and ``muxbleed'' associated with
well-exposed pixels in IRAC channels 1 (3.6 $\mu$m) and 2 (4.5 $\mu$m).
The muxbleed corrections were empirically derived from a large
set of IRAC data taken with the same exposure time.  We then reran the
MOPEX mosaicing software with modified cosmic-ray rejection parameters and
an output pixel scale reduced by 50\%\ to improve the resolution of the
final images.  The spatial resolution in the combined mosaics is $\sim
2\arcsec$ FWHM.  The $1 \sigma$ sensitivity of our final mosaics was
measured by calculating the standard deviation between the flux values
in 1\farcs8 radius
beams placed along random blank points in the mosaic.  The resultant
1$\sigma$ point source detection limits, corrected for beam size, were 0.4, 0.3, 0.9 and 
1.0~$\mu$Jy at 3.6, 4.5, 5.8 and
8.0~$\mu$m, respectively.  The expected $1 \sigma$ background-limited
point source sensitivity values are 0.14, 0.23, 1.85 and 1.93~$\mu$Jy 
for medium background.
The difference between the two sets of values, especially
at the shortest wavelengths,
is mostly attributable to confusion noise and partly due to the correlated
noise between pixels.

The $z=6.56$ source (HCM~6A) is detected in the 3.6 and
4.5~$\mu$m mosaics (Figure~1).  At the spatial resolution of IRAC, the
neighboring, bright galaxy $\sim$6\arcsec\ to the southwest ($z = 0.375$) could
be contaminating the photometry of the high-redshift source.  We first
attempted to perform the photometry on HCM~6A using a priors-based
deblending technique whereby we use the position of sources in 
high resolution {\it Hubble Space Telescope} WFPC2 images of the field, convolve them
with the {\it Spitzer} point spread function (PSF) and subtract them
out. However, because of the extended morphology of the foreground galaxy,
likely morphological $k$-corrections, and the position dependence of
the IRAC PSF, the residuals from this technique were found to be rather
large and unsuitable.  Prior-based deblending is most effective when
sources are unresolved.

We then adopted an alternate approach.  After measuring aperture-corrected
fluxes of $0.64 \pm 0.14\, \mu$Jy and $1.42 \pm 0.28\, \mu$Jy at 3.6
and 4.5~$\mu$m, respectively, for HCM~6A, we attempted to estimate
the contribution of the foreground galaxy to these values.  We derived
surface brightness profiles of the bright galaxy in expanding annuli,
fit those with an analytical function and extrapolated the function to
assess the surface brightness of the galaxy at the location of HCM~6A. We
find that the the maximal contribution from the galaxy to the flux of
HCM~6A is $0.1 - 0.2\, \mu$Jy at 3.6~$\mu$m and 4.5~$\mu$m
and 1.7~$\mu$Jy at 8.0~$\mu$m, depending on the functional form of
the surface brightness profile. We subtract an average value of $0.14\, \mu$Jy
and $0.17\, \mu$Jy at 3.6 and 4.5~$\mu$m, respectively, for the galaxy 
contribution and adopt flux values of
$0.5 \pm 0.2\, \mu$Jy and $1.25 \pm 0.3\, \mu$Jy at 3.6 and 4.5~$\mu$m, respectively,
for HCM 6A. 

At 5.8~$\mu$m we provide a $1 \sigma$ upper limit of 0.9~$\mu$Jy
while at 8.0~$\mu$m there is a hint of a positive flux, $0.8 \pm 1.0~\mu$Jy. 
Given the noise properties of the mosaic and blending from the
foreground, bright galaxy, we classify the 5.8~$\mu$m and 8.0~$\mu$m
values as 1$\sigma$ upper limits, the 3.6~$\mu$m value as a $\sim3\sigma$
detection, and the 4.5~$\mu$m value as a $\sim 5 \sigma$ detection. The
faintness of the source and detectability at the limits of the IRAC
instrument, despite a lensing amplification factor of 4.5, illustrate the
difficulty associated with a detailed study of the $z>6$ galaxy population.

To assess the reliability of the source and the photometry, we selected
a random 2/3 of the 4.5$\mu$m frames and re-generated the mosaic, repeating the process
3 times. The photometry was performed on each of these mosaics, with
the difference between the values yielding a systematic uncertainty which
is factored into the aforementioned flux density estimates. We also
investigated the effect of modifying the mosaicing parameters, which included
making the outlier rejection threshold more stringent. We find the photometry
to be insensitive to these parameters.

Finally, we briefly note that the peak of the 4.5~$\mu$m emission is
offset mostly to the north from the peak of the 3.6~$\mu$m emission by
$1\farcs1 \pm 0\farcs2$.  This is less than the FWHM of the PSF at these
wavelengths, but is much greater than the 0\farcs2 relative astrometric
accuracy between the IRAC images in the various channels, as measured
from the scatter in the centroid of bright sources.  
Furthermore, the peak of the 3.6~$\mu$m emission is well aligned with the continuum
source detected by \citet{Hu:02}. This offset
is most likely attributable to the extended nature of the Ly-$\alpha$
source. It is possible that one of the two optical components of HCM 6A
is dominating the
H$\alpha$ emission, affecting the 4.5~$\mu$m morphology and centroiding.
Regardless, since 
the separation of the two components is less than the IRAC PSF, we only 
consider emission from the entire ensemble.

\section{Interpretation}

Figure~2 shows the spectral energy distribution (SED) of HCM~6A at all
detected wavelengths including the ground-based photometry
from \citet{Hu:02}. Since the Balmer break for a galaxy at $z=6.56$
falls shortward of the IRAC 3.6~$\mu$m filter and the source is known
to have a strong Ly$\alpha$ emission line, we interpret the upturn in
the SED at 4.5~$\mu$m flux to be due to the presence of H$\alpha$ which,
at a redshifted wavelength of 4.96~$\mu$m,
falls within the IRAC 4.5~$\mu$m passband. The presence of Ly$\alpha$
suggests that the object is a young star-forming galaxy. We tentatively
assume that the continuum follows
a simple $f_\nu \propto \lambda^{0}$ spectrum and fit it to the flux density at
wavelengths which are not contaminated by line emission, i.e. $J$,
$H$, $K$ and 3.6~$\mu$m.  Extrapolating this continuum to 4.5~$\mu$m
yields a continuum level of $0.38\, \mu$Jy.  The continuum is slightly
decreased to $0.33\, \mu$Jy if the $z$-band photometric point is included
but can be as high as $0.7\, \mu$Jy if the continuum slope is red due
to dust or an evolved stellar population. These estimates are about a factor of 2$-$3 below the measured flux of
$1.25\, \mu$Jy at 4.5~$\mu$m.
We attribute the difference to the presence of H$\alpha$+[\ion{N}{2}]
emission.  The resultant combined line luminosity is in the range
$(3-15) \times 10^{43}$~erg~s$^{-1}$ which factors in the uncertainty
in the continuum level as well as the photometry.  When corrected for the lensing magnification
factor of 4.5 and a canonical [\ion{N}{2}]/H$\alpha$ ratio of 0.3
\citep{Tresse:99}, this implies a H$\alpha$ line luminosity of $(5-25)
\times 10^{42}$~erg~s$^{-1}$ and a resultant star-formation rate of
140$\pm$90~M$_{\sun}$ yr$^{-1}$ using the \citet{Kennicutt:83} SFR relation.
The implied rest-frame H$\alpha$
equivalent width of $\sim$0.2~$\mu$m
is not unreasonable for young stellar populations
\citep[c.f.,][]{Leitherer:95b}.
Typical systematic uncertainties are about a factor of 4, arising
from the uncertainty in the lensing magnification factor ($\times$2) and the 
[\ion{N}{2}]/H$\alpha$ ratio which can span the 0.1-1 range.
The derived SFR is about an order of magnitude higher than that
derived from the UV continuum \citep{Hu:02}. Assuming this is a result
of dust, the implied extinction is $A_{\rm V}\sim$1.0~mag. If this
interpretation is correct, the presence of dust 
is remarkable, requiring a significantly evolved stellar population
to already be in place by redshift $z = 6.56$.  Apparently, even at
the highest redshifts currently probed, we are still not identifying
primordial stellar populations.  This is consistent with spectroscopic
studies of the most distant galaxies.  \ion{He}{2} $\lambda$1640~\AA,
predicted to be a major coolant in very low metallicity systems and thus
a good tracer of Population~III stars, remains undetected in stacked
spectra of multiple, high-redshift galaxies \citep{Dawson:04} or in
ultradeep spectra of individual galaxies \citep{Nagao:05}.  The likely
implication of this dust detection and the \ion{He}{2} non-detections is
that significant star formation occurs at redshifts $> 10$.

We next attempted to fit the photometric measurements of the galaxy with a
\citet{Bruzual:03} template SED to constrain its mass.  
Templates with metallicity values of 0.02 solar and with a Salpeter IMF were adopted.
The IMF at these redshifts is a significant source of uncertainty since a top-heavy IMF
could produce the same emission line flux with a considerably lower SFR.
The lensing factor
of 4.5 was removed from the flux estimates. The uncertainty in the lensing
magnification factor does not dominate
the uncertainties in the SED fitting which are driven by the large photometric
errors. 
The 4.5~$\mu$m photometry
was replaced by the continuum level after the H$\alpha$+[\ion{N}{2}]
lines were subtracted. We find that the best fit template corresponds
to a stellar mass of $8.4 \times 10^{8}$~M$_{\sun}$, an instantaneous
starburst age of 5 Myr, and $A_{\rm V} = 1.0$~mag.  The low resultant
age is consistent with the strong emission lines in this galaxy although
the mass estimate is uncertain to $\times$10.

An alternate interpretation is that the flux in all passbands is dominated
by the continuum emission rather than the line emission. Repeating the
Bruzual-Charlot fits results in a galaxy with mass in stars of $4.0 \times
10^{9}$~M$_{\sun}$, a starburst age of 30 Myr and $A_{\rm V} = 0.7$~mag.
Template
fits with extinction forced to zero, ages of 700 Myr and masses of $7 \times
10^{9}$~M$_{\sun}$ also provide reasonable fits to the photometry (Figure 2).
However, the formal, reduced $\chi^{2}$ estimate for these fits is 
about 2$-$3 times higher than fits with the H$\alpha$ contribution subtracted.

The remarkable agreement between the extinction values independently derived from
the ratio of H$\alpha$ to Ly$\alpha$ emission lines and the
fit to the broadband photometry, therefore
leads us to conclude that our first hypothesis, that
the 4.5~$\mu$m flux is dominated by H$\alpha$ in emission and that the
galaxy has an intrinsic extinction of $\sim 1.0$~mag, is more amenable
to the data.

\section{Discussion}

We assess the possibility that the unusual photometry of this narrow-line
Ly$\alpha$ emitter (LAE) could be due to the presence of an AGN. 
A $\sim$100ks {\it Chandra} X-ray observation of this
field does not reveal the presence of an X-ray counterpart for HCM 6A.
Based on the data presented in \citet{Barger:01b} and \citet{Bautz:00},
the lensing uncorrected
limiting flux is $\sim$10$^{-15}$~erg~cm$^{-2}$~s$^{-1}$ between 2-7 keV 
and 3$\times$10$^{-16}$~erg~cm$^{-2}$~s$^{-1}$ between 0.5-2 keV. 
This translates to F$_{\rm X}$/F$_{\rm opt}<0.06$ which is small but
still more than a factor of 5
larger than the range of flux ratios seen for the Sloan quasars at
these redshifts \citep{Brandt:02}. The Ly$\alpha$ to X-ray flux ratio
limit for HCM6A is $>0.04$. 
This limit is within the range of high redshift Type II QSOs.
However, deep X-ray observations
of the narrow-line Ly$\alpha$ emitter population at $z\sim4.5$ yield similar
detection limits for individual objects. X-ray stacking of these 101 LAEs
yields a non-detection which is $\sim20$ times higher than the ratio
for Type II QSOs implying that less than 5\% of these objects could have AGN
\citep{Wang:04}. Since HCM6A is very similar in properties
to the LAEs, we infer that it too does not host an AGN.


In Figure~3, we compare the photometric properties of HCM~6A discussed
in this paper with other $z > 6$ candidates from \citet{Egami:05}
and \citet{Mobasher:05}. 
HCM 6A is the only one of these sources with detected Ly$\alpha$ and with
a spectroscopically confirmed redshift. The derived ages of the stellar population for
the other sources are much larger, spanning $30 - 1000$~ Myr, while HCM~6A
spans $5 - 30$~Myr, albeit with large uncertainties due to the
lower flux density and correspondingly lower signal to noise. The mass
derived for the \citet{Egami:05} galaxy is very similar to HCM~6A while
the estimate of the mass for the \citet{Mobasher:05} galaxy is three
orders of magnitude larger. 
The range of masses and ages spanned at $z\sim6.5$ suggests that the growth
of galaxies occurs along different evolutionary channels,
whereby some galaxies undergo monolithic collapse,
forming the bulk of their stars in a single burst at very high redshift,
while others, like HCM~6A, build their stellar mass in the hierarchical 
merging scenario.

If HCM~6A is typical of the emission line galaxy population as seems to be
the case, reddening
could explain the sharp decrease in the cosmic star-formation rate
seen at $z \sim 6$ by \citet{Stanway:04}, \citet{Bouwens:04} and
\citet{Taniguchi:05} relative to the extinction corrected values
of \citet{Giavalisco:04b}.  HCM~6A has a derived dust extinction of
$A_{\rm V} = 1.0$~mag, corresponding to $A_{\rm UV}$ of 2.6~mag.
This would boost the $z > 6$ star-formation rate of $5.7 \times
10^{-4}$~M$_{\sun}$~yr$^{-1}$~Mpc$^{-3}$ derived by \citet{Taniguchi:05} by an order of magnitude.
Dust is predominantly formed in the shells of red giant
stars which would barely have sufficient time to evolve when the age
of the Universe is 850~Myr.  The implication is that the first epoch of
star-formation in HCM 6A,
occured at $z \sim 20$. Furthermore, if dust reddening were an unexpectedly
significant factor at $z>6$, the re-ionization of the
IGM could be accounted for by the luminous end of the
star-forming galaxy population without resorting to a steep faint end slope.
Measures of dust extinction in a larger sample of emission-line galaxies
can test this hypothesis and is currently under investigation.

\acknowledgements 

We are very grateful to Esther Hu for generously providing images
of HCM6A which enabled us to check astrometric alignment.
We also wish to thank Mark Bautz for kindly estimating the X-ray upper limits
from the {\it Chandra} observation.
This work is based on observations made with the {\it Spitzer Space
Telescope}, which is operated by the Jet Propulsion Laboratory, California
Institute of Technology, under a NASA contract.  Support was provided
by NASA through an award issued by JPL/Caltech.

\begin{figure}
\plotone{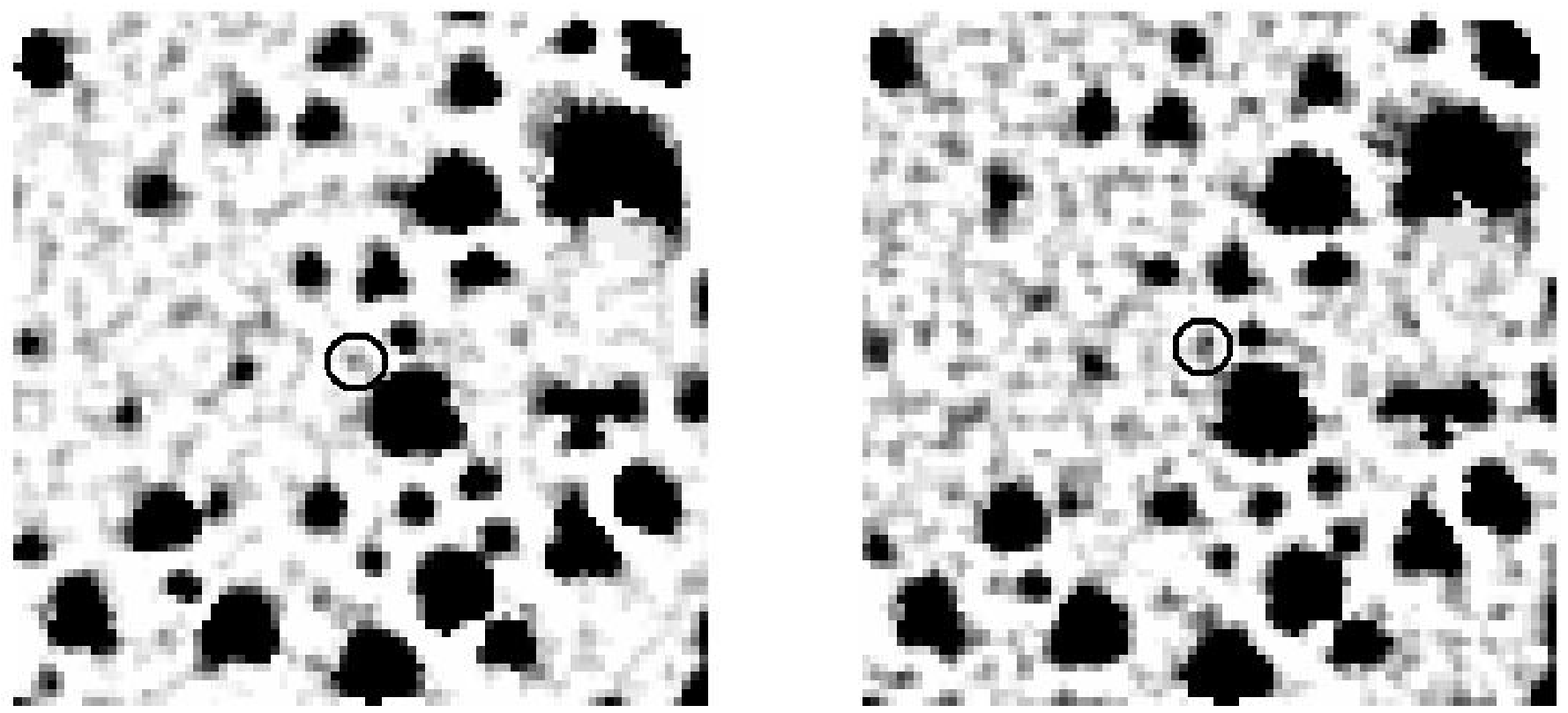}

\caption{ 3.6~$\mu$m (left) and 4.5~$\mu$m (right) snapshots of the
$z=6.56$ emission line object HCM~6A (circled), lensed by Abell~370.
The snapshots are 48$\arcsec$ on a side with North up and East to
the left, and are centered on the location of HCM~6A, $\alpha = 2^{\rm
h}39^{\rm m}54.73^{\rm s}, \delta = -1\deg33\arcmin32\farcs3$ (J2000). The
figure has been generated by median filtering the real mosaic with a $6
\times 6$ pixel ($3\farcs6 \times 3\farcs6$) boxcar and subtracting that
from the mosaic. This has the effect of removing any smooth, underlying
structure and amplifying the compact sources, thereby clearly separating
HCM~6A from the $z=0.375$ foreground cluster galaxy to the south-west.}

\end{figure}

\begin{figure}
\plotone{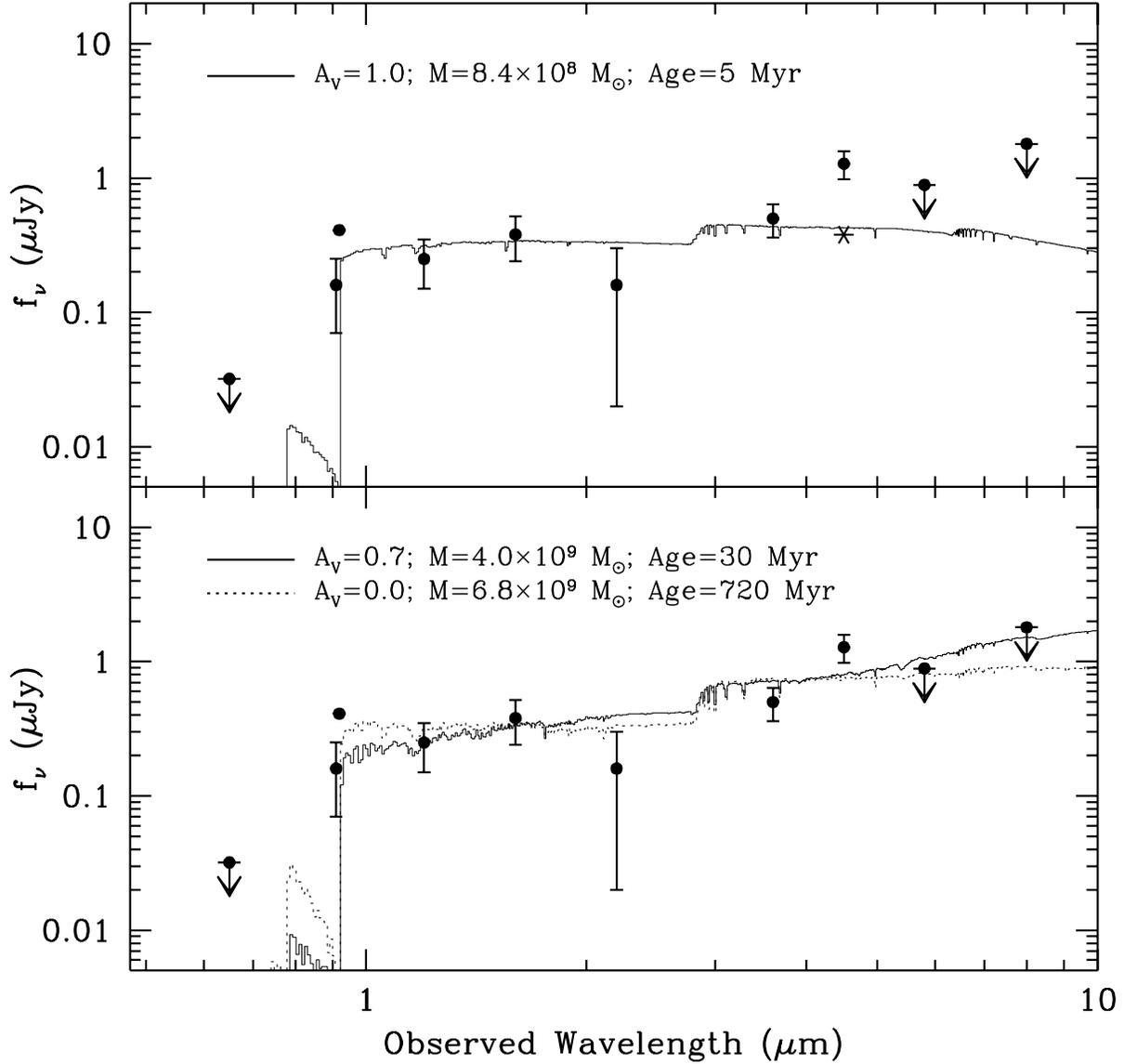}

\caption{ Published ground-based optical photometry of the $z=6.56$ galaxy
HCM~6A \citep{Hu:02} shown along with the {\it Spitzer}/IRAC photometry
presented in this paper. Also shown is a best fit Bruzual-Charlot
template spectral energy distribution.  In the upper panel, the broadband
photometry at 4.5~$\mu$m has been corrected for the estimated strength
of the H$\alpha$ emission line down to $0.38\, \mu$Jy (asterisk) before
the fit shown by the solid line was performed. In the lower panel, the
4.5~$\mu$m photometry was assumed to be dominated by the continuum. The
extinction in magnitudes, stellar mass in solar masses and age of the
stellar population for the templates are shown in the panels. The masses
shown have been corrected for the lensing amplification factor of 4.5,
though the photometry has not been.}

\end{figure}

\begin{figure}
\epsscale{0.8}
\plotone{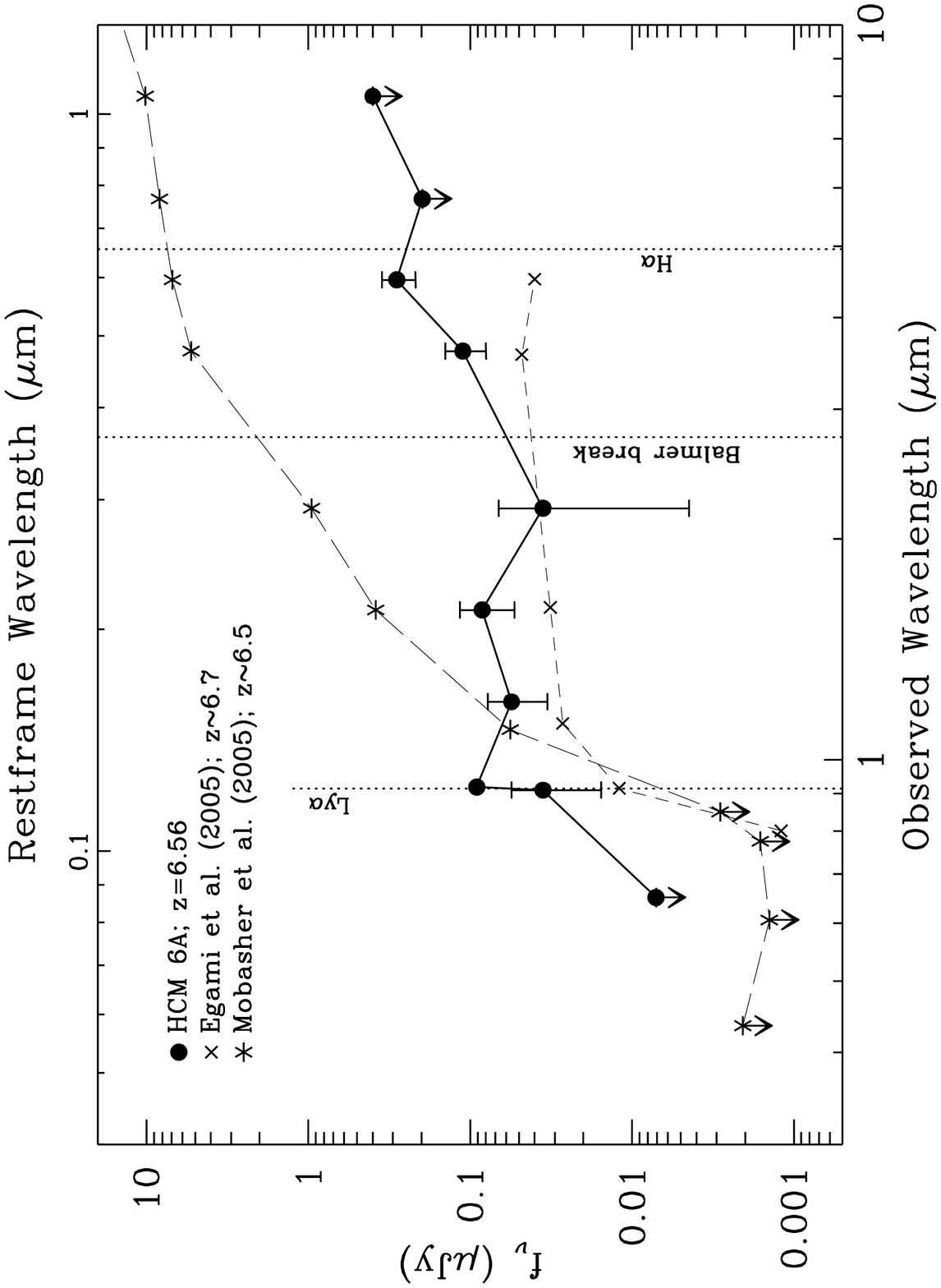}

\caption{Comparison between the properties of the $z=6.56$ object
discussed in this paper with other $z>6$ candidates from \citet{Egami:05}
and \citet{Mobasher:05}.  The photometry of HCM~6A has been corrected
by the lensing amplification factor of 4.5 and the photometry of the
\citet{Egami:05} source has been corrected by the lensing amplification
factor of 25.  HCM~6A is the only one of the three galaxies which has
been spectroscopically confirmed to reside at $z \sim 6.5$.  The top
axis and vertical dotted lines refer correspond to $z = 6.56$.}

\end{figure}

\end{document}